\documentclass{emulateapj}
\usepackage{apjfonts}
\usepackage{lscape}

\newcommand{\mum}{\ifmmode{\rm \mu m}\else{$\mu$m}\fi}

\newcommand{\spitzer}{{\em Spitzer}\ }

\begin{document}


\title{Probing the excitation of extreme starbursts: High resolution
mid-IR spectroscopy of Blue Compact Dwarfs}

\author{
Lei~Hao \altaffilmark{1},
Yanling~Wu\altaffilmark{2},
V.~Charmandaris\altaffilmark{3,4},
H.~W.~W.~Spoon\altaffilmark{5},
J.~Bernard-Salas\altaffilmark{5},
D.~Devost\altaffilmark{6},
V.~Lebouteiller\altaffilmark{5},
J.~R.~Houck\altaffilmark{5}
\email{haol@astro.as.utexas.edu}
}
\altaffiltext{1}{University of Texas at Austin, McDonald Observatory, 1 University Station, C1402, Austin, Texas 78712-0259}
\altaffiltext{2}{Caltech, Infrared Processing and Analysis Center, MC 314-6, Pasadena, CA 91125}
\altaffiltext{3}{University of Crete, Department of Physics,
  P.~O. Box 2208, GR-71003, Heraklion, Greece}
\altaffiltext{4}{IESL/Foundation for Research \& Technology - Hellas,
  GR-71110, Heraklion, Greece, and Chercheur Associ\'e, Observatoire de
  Paris, F-75014, Paris, France}
\altaffiltext{5}{Cornell University, Astronomy Department,   Ithaca, NY 14853-6801}
\altaffiltext{6}{CFHT:Canada France Hawaii Telescope, Kamuela, HI, 96743, USA.}

\begin{abstract}

  We present an analysis of the mid-infrared emission lines for a sample of
  12 low metallicity Blue Compact Dwarf (BCD) galaxies based on high
  resolution observations obtained with Infrared Spectrograph on board
  the \spitzer Space Telescope. We compare our sample with a local
  sample of typical starburst galaxies and active galactic nuclei
  (AGNs), to study the ionization field of starbursts over a broad
  range of physical parameters and examine its difference from the one
  produced by AGN. The high-ionization line [OIV]25.89$\mu$m is
  detected in most of the BCDs, starbursts, and AGNs in our sample. We
  propose a diagnostic diagram of the line ratios
  [OIV]25.89$\mu$m/[SIII]33.48$\mu$m as a function of
  [NeIII]15.56$\mu$m/[NeII]12.81$\mu$m which can be useful in
  identifying the principal excitation mechanism in a galaxy. Galaxies
  in this diagram split naturally into two branches. Classic AGNs as
  well as starburst galaxies with an AGN component populate the upper
  branch, with stronger AGNs displaying higher [NeIII]/[NeII]
  ratios. BCDs and pure starbursts are located in the lower branch. We
  find that overall the placement of galaxies on this diagram
  correlates well with their corresponding locations in the
  log([NII]/H$\alpha$) vs. log([OIII]/H$\beta$) diagnostic diagram,
  which has been widely used in the optical. The two diagrams provide
  consistent classifications of the excitation mechanism in a
  galaxy. On the other hand, the diagram of
  [NeIII]15.56$\mu$m/[NeII]12.81$\mu$m
  vs. [SIV]10.51$\mu$m/[SIII]18.71$\mu$m is not as efficient in
  separating AGNs from BCDs and pure starbursts.  Our analysis
  demonstrates that BCDs in general do display higher [NeIII]/[NeII]
  and [SIV]/[SIII] line ratios than starbursts, with some reaching
  values even higher than those found at the centers of AGNs. Despite
  their hard radiation field though, no [NeV]14.32$\mu$m emission has been
  detected in the BCDs of our sample.

\end{abstract}




\section{Introduction}

Starburst galaxies are hosts of intense star formation activity with
star formation rates (SFR) more than a factor of 3 compared to the
Milky Way (SFR$>$10M$_{\odot}$yr$^{-1}$). The stellar population of
the starbursts, especially the number of massive stars and the slope
of the initial mass function (IMF), directly affect the hardness of
the ionization field in the interstellar medium (ISM). Therefore,
ratios of emission lines of different excitation levels are powerful
tools to study the properties of the ionization field and to infer the
stellar population of the starbursts. Infrared emission-line ratios
are of particular interest and have been explored extensively with the
Infrared Space Observatory \citep[i.e.][]{tho_etal_00,for_etal_01,
  ver_etal_03, ver_etal_05}, as
the active star forming regions in the disks of starburst galaxies and
circumnuclear regions are often optically obscured by dust. As a
result, the emitted radiation, originating from a massive starburst
and/or an accreting active nucleus, is emerging in the mid-infrared
(mid-IR) \citep[see][]{vac_etal_02, gor_etal_01} and far-IR. The
advent of the Spitzer Space Telescope with its improved sensitivity
opens new avenues in this direction, enabling detailed studies of
large samples of quiescent and star forming galaxies \citep{dal_etal_06, dal_etal_09, oha_etal_08, ber_etal_09}, as
well as more complex dust enshrouded infrared luminous sources which
often host an active galactic nucleus (AGN) \citep{arm_etal_07, far_etal_07}.

Estimating the hardness of the ionization field plays a key role in
determining the dominant power source at the centers of
galaxies. Nuclear starbursts and accreting super-massive black holes
(SMBH) are often intertwined and coexist, as best demonstrated in the
case of Ultra Luminous Infrared Galaxies (ULIRGs)\citep[see e.g.][and
references therein]{gen_etal_98, arm_etal_07}. The radiation of an
AGN, produced from the accretion of gas and stars onto the SMBH, has a
much harder spectrum compared to the one produced by the massive young
stellar populations. Therefore, the hardness of the radiation field
has been a key parameter addressed by previous diagnostic methods to
determine the energy source in galaxies \citep[e.g.][]{bal_etal_81,
  vei_ost_87, gen_etal_98, lau_etal_00, nar_etal_08}.

In the study of the ionization field of galaxies with massive
starbursts, the case of low-metallicity Blue Compact Dwarfs (BCDs) is
particularly interesting. BCDs are galaxies characterized by their
blue optical colors (M$_{\rm B}$$<$-18\,mag), small linear sizes ($<$1kpc)
and low luminosities \citep{thu_mar_81, gil_etal_03}. Even though BCDs
are defined based on their morphological properties, they have been
ubiquitously found to have low oxygen abundances as measured from
the ionized gas in HII regions. 

It is well established that the hardness of the stellar radiation
increases with decreasing metallicity \citep[e.g.][]{cam_etal_86}. As
a result, the ionization field of BCDs is generally much harder
compared to a typical starburst. It is also known that the mass of a
super massive black hole (SMBH) is correlated with the mass of the
bulge of the host galaxy \citep{mag_etal_98, geb_etal_00, fer_mer_00} while only a
fraction of the most massive SMBH are actively emitting radiation as a
result of mass accretion \citep{hao_etal_05_agn2}. BCDs in general are not massive enough to
host such an active SMBH, with possibly only a few exceptions \citep[see][]{izo_etal_07, izo_thu_08}.  Consequently, the excitation mechanism for the hard radiation
field in BCDs is likely significantly different from the one
responsible for the radiation field produced by an AGN. Furthermore,
unlike starburst galaxies which are often obscured by dust, an
analysis of a large sample of BCDs suggests that in BCDs nearly always
the young massive stars which ionize the interstellar medium are
easily detected and their dust extinction measured from the optical Balmer lines is typically fairly low \citep{wu_etal_08b}.  There might be exceptions indicated from the mid-IR observations, with SBS0335-052E as the most extreme case. However, this is rather a unique galaxy and deviates from
the usual trends seen in BCDs for different reasons as discussed in
detail by \citet{hou_etal_04_sbs}. Understanding the differences in
the excitation mechanism between BCDs and AGNs can provide insightful
clues to decipher the power source in a starburst galaxy also hosting
an AGN. Identifying those differences from the mid-IR spectral
features is one of the main motivations behind the present work.

BCDs are often much fainter than typical starburst galaxies in the
infrared. Up until the launch of the \spitzer Space Telescope
\citep{wer_etal_04}, studies of starburst galaxies in the IR have
included only a handful of BCDs \citep{lut_etal_98, tho_etal_00,
  ver_etal_03, mad_etal_06}. The superb sensitivity of the \spitzer
allows us to observe a large sample of BCDs \citep[see][]{wu_etal_06,
  wu_etal_08}. In this paper, we provide the ionized gas diagnostic
analysis of the low metallicity BCDs accessible to the high-resolution
modules of the Infrared Spectrograph \citep[IRS,][]{hou_etal_04}. In
addition, we contrast the BCD sample with typical starbursts and AGNs,
to probe the excitation mechanism in these types of objects and
understand where the difference comes from.

In \S 2, we introduce the sample and the data reduction method. In \S
3, we study the high-ionization emission lines, and propose a new
diagnostic diagram to determine the energy source in a galaxy. In \S
4, we focus on other infrared emission lines and compare various
excitation diagnostic diagrams of starbursts and BCDs. Our findings
are summarized in \S 5.


\section{Sample and Data Reduction}

\subsection{The BCD sample}

We have compiled a sample of 12 BCDs and observed them in starring
mode with the high-resolution modules (short-high, SH:
4.7$\arcsec$$\times$11.3$\arcsec$ and long-high, LH:
11.1$\arcsec$$\times$22.3$\arcsec$) of the IRS as part of the
instrument team's Guaranteed Time Observation (GTO) programs. The
galaxies were drawn from the sample of \citet{wu_etal_06}. We exclude
Tol1214-277 as it is too faint and its high-resolution spectrum has
low signal-to-noise ratio. We also exclude CG0752, CG0563, and CG0598
as they do not show typical BCD signatures. The characteristics of
individual objects in the sample and analysis based on their \spitzer
low-resolution spectra are presented in \citet{wu_etal_06, wu_etal_08,
  wu_etal_08b}.

The high-resolution module slits were centered on the nucleus of each
galaxy, as was the case of the low-resolution observations
\citep{wu_etal_06}. The data were processed by the \spitzer Science
Center (SSC) data reduction pipeline version 13.2.  Individual
pointings to each nod position of the slit were co-added.  The median
of the combined images were extracted using the full slit extraction
method of the {\em IRS} data analysis package SMART
\citep{hig_etal_04}. Then the extracted spectra were flux calibrated
by multiplying them with a relative spectral response function (RSRF),
which was created from the {\em IRS} standard star, $\xi$ Dra, for
which accurate templates were available \citep{coh_etal_03}.  Note
that emission from the sky background has not been subtracted from our
spectra due to the lack of dedicated sky observations, thus
discontinuity exists in some source spectra between SH and LH,
however, this will not affect the results of our analysis since our
conclusions are based on ratios of emission lines observed in the same
module.

In the Appendix, we include close up plots of strong and weak emission
lines visible in the IRS high-resolution spectra of BCDs. In order to
measure the line strengths, a Gaussian profile is fitted to the
lines above a local continuum fitted by an one-order polynomial. The [OIV]25.89$\mu$m and [FeII]25.99$\mu$m are close to each
other and they are fit together using a two-Gaussian profile with
fixed rest-frame wavelengths. The measured line strengths for the BCD
sample are listed in Table~\ref{tabline}. The errors are estimated
based on the error propagation of the Gaussian fit assuming the error of the flux as the root mean square ($rms$) of the continuum fit. Note that this does not take into account all the systematic uncertainties, especially the flux calibration. We estimate these to be 5\% and choose it as our minimum uncertainty. The upper-limits are obtained by measuring the flux of a Gaussian with a height three times the local $rms$, and with a FWHM equal to the instrumental resolution.

No extinction correction has been performed in our analysis. The
corrections are typically minor for mid-IR fine-structure line ratios,
except for ratios involving lines found close to the silicate
features, such as the [SIV]10.51$\mu$m line. In the analysis of
diagrams involving the [SIV] line (see Figure~\ref{sratio_neratio}),
we provide arrows indicating the effects of an A$_{\rm V}$=30 mag of
extinction, assuming a uniform dust screen and a \citet{li_dra_01}
extinction curve \citep[see][]{far_etal_07}.

\subsection{The Starburst and AGN sample}

In order to study the ionization field of starbursts over a broad
range of physical properties, we also examine a sample of 24 local
starbursts observed in staring mode with the IRS high-resolution
modules. The details of the sample, their high-resolution spectra and
line measurements are described in \citet{ber_etal_09}. It should be
emphasized that even though we broadly use the term ``starburst'' when
referring to them, some may also have a weak AGN component.

In addition, we created a comparison sample of typical AGNs to investigate how
the difference in the excitation mechanism of the gas, namely the
radiation produced by an accretion disk versus massive young star
formation, is reflected in the mid-IR emission lines. It is not our
intention to have an AGN sample that is complete, but rather one that
is representative of typical AGNs. Hence, we choose bona-fide AGNs,
with no ambiguity in their classification, where the nuclear emission
is known to dominate the integrated spectra of their host galaxies. We
select from the Type 1 sources (Quasars and Seyfert 1s) used in
\citet{hao_etal_07} for which IRS high-resolution observations are
available. We also examine the strength of the Polycyclic Aromatic
Hydrocarbon (PAH) emission in their mid-IR spectra and exclude sources
with strong PAH features, since this would suggest a significant
contribution from star forming regions in the disk of the host
galaxy. Among those sources with weak PAH emission, we choose 10 AGNs
that show strong mid-IR emission lines, in order to obtain the best
S/N possible in our measurements. The AGN IRS spectra were extracted
and the emission lines measured in the same way as the BCDs (see \S
2.1).

Furthermore, whenever applicable, we also compare the properties of
emission lines in BCDs with those in low-redshift ULIRGs as well as
HII regions found in our Galaxy, and in the Large and Small Magellanic
Clouds (LMC/SMC). The sample and line measurements of the low-redshift
ULIRGs are taken from \citet{far_etal_07}, and the HII region data
from \citet{pee_etal_02} and \citet{leb_etal_08}.

%

\section{High-Ionization Emission Lines Diagnostics}
\subsection{The [OIV]25.89\mum\  Emission}

The IRS high-resolution spectra of BCDs are rich in fine-structure
emission lines. All BCDs display strong [NeIII]15.56$\mu$m,
[SIII]18.71$\mu$m, and [SIV]10.51$\mu$m emission. The [NeII]12.81$\mu$m
line is well detected in about half of the BCD sample, but it is not
seen in SBS0335-052E, Tol65, and VIIZw403. In the comparison samples
of starbursts and AGNs, all these lines are strong and we discuss their
relative flux ratios in \S 4.

The presence of high-ionization emission lines is of great interest,
since their excitation often requires extreme conditions. One
particular case is the [OIV]25.89$\mu$m line which has an ionization
potential (IP) of 54eV, just above the HeII edge, and therefore it
cannot be readily generated in the HII regions surrounding
main-sequence stars. However, the [OIV] line had already been
detected in several starburst galaxies with ISO
\citep[see][]{lut_etal_98}. A possible explanation for this was that
the [OIV] emission is due to a dust enshrouded AGN as the hard
radiation field produced by the AGN can easily provide the energy
needed to ionize O$^{3+}$. However, \citet{lut_etal_98} showed that at
least in some starburst systems (such as the outer regions of M82),
the [OIV] cannot be attributed to a weak AGN. Instead it must be due
to either very hot stars, such as the Wolf-Rayet (WR) stars \citep[see
also][]{schaer_sta_99} or to the ionizing shocks associated with the
starburst activity.

With the superb sensitivity of the \spitzer Space Telescope, we now
find that the [OIV] line is fairly common in our BCD sample as well as
the comparison sample of starburst galaxies. We detected [OIV]
emission in 8 of the 12 BCDs, and in 21 of the 24 starbursts
\citep[see][]{ber_etal_09}.  This is very interesting since, as we
will discuss in the following sessions, line ratios involving [OIV]
provide a good diagnostic tool for identifying the presence of an AGN
\citep[see also][]{gen_etal_98, arm_etal_07}.

\subsection{The [OIV]/[SIII] vs. [NeIII]/[NeII] diagram}

\begin{figure}[ht]
\centerline{
\includegraphics[angle=0.,width=\hsize]{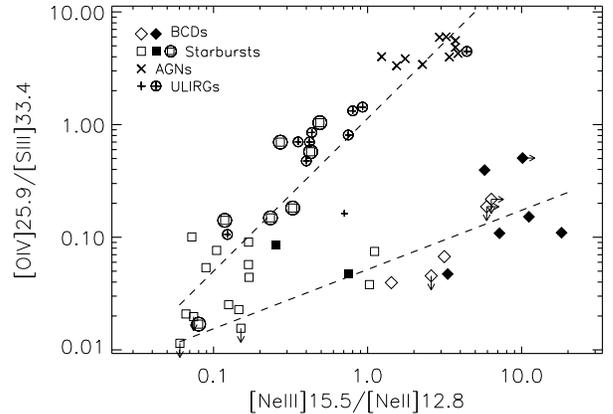}}
\caption{A plot of the [OIV]25.89$\mu$m/[SIII]33.48$\mu$m as a function of [NeIII]15.56$\mu$m/[NeII]12.81$\mu$m, for BCDs (diamonds),
  starbursts (squares), AGNs (crosses), and ULIRGs (plus signs). A circle
  around a symbol indicates that  [NeV]14.32$\mu$m emission has been detected,
  while a filled symbol implies that this galaxy also has been classified as a Wolf-Rayet (WR) in the optical (see \S 3.4). The
  error bars of the line ratios in this and all subsequent plots are
  similar to the symbol size. The galaxies appear to be localized on
  two branches. One branch consists mainly of AGNs, or
  starbursts/ULIRGs hosting an AGN, while the other mainly contains
  BCDs or pure starburst galaxies. Two dashed lines are drawn as a
  guide to identify the two branches.  
  \label{os3_neratio}}
\end{figure}

In Figure~\ref{os3_neratio}, we plot the line ratios of
[OIV]25.89$\mu$m/[SIII]33.48$\mu$m as a function of
[NeIII]15.56$\mu$m/[NeII]12.81$\mu$m for our BCDs, starburst galaxies
 AGNs and ULIRGs. We choose to use the line ratio of [OIV]25.89$\mu$m and
[SIII]33.48$\mu$m because of their large difference in the ionization
potential, and the absence of aperture effects, as both lines reside
in the same IRS long-high module (see also Figure 3 in
\citealt{gen_etal_98}) . We use the [NeIII]/[NeII] line ratio as the
other variable, as it is commonly used as an indicator of the strength
of the EUV radiation field \citep[e.g.][]{lut_etal_98, tho_etal_00,
  ver_etal_03}. Since both [NeIII] and [NeII] lines are of the same
element, the ratio is independent of the gas abundance. We clearly see
that the [OIV]/[SIII] ratio separates AGNs from BCDs, with AGNs having
in general a factor of 10 higher [OIV]/[SIII] ratios than BCDs, even
though they have similar [NeIII]/[NeII] ratios. Starburst galaxies
display smaller [NeIII]/[NeII] ratios than AGNs and BCDs, but their
values of the [OIV]/[SIII] ratio vary. In particular, it appears that
starburst galaxies populate two branches with a common origin (the
dashed lines in Figure~\ref{os3_neratio}), one pointing towards the
locus of AGNs and the other to the BCDs. Among the starburst galaxies,
Mrk266, NGC1365, NGC2623, NGC4194, NGC4945, NGC520, NGC660, NGC253,
NGC1614, NGC1097, NGC3079, NGC4676, and NGC7252 are located in the
upper branch reaching to the AGN locus, while NGC1222, NGC2146,
NGC3256, NGC3310, NGC3628, NGC4088, and NGC7714 are in the lower
branch that ties to the BCDs. The galaxies IC342, NGC3556, and NGC4818
have no detected [OIV] emission and their upper limits are indicated
with arrows on the plot.

Given that the gas in AGNs and BCDs is excited by distinctly different
physical mechanisms, the fact that they are found at the corresponding
ends of the two branches of the diagram, also suggests that the
excitation mechanisms leading to the [OIV] line emission in starbursts
galaxies populating the two branches is different. This could be
understood if the mid-IR emission of starbursts in the upper branch is
due to some AGN component, while in starbursts of the lower branch is
only due to the reprocessing of pure stellar radiation.  Therefore,
such a diagram can be used as a diagnostic tool to determine the
origin of the dust enshrouded energy source in star forming
galaxies. Interestingly, among the starbursts located in the upper
branch, six also have detected [NeV]14.32$\mu$m emission
\citep[see][]{ber_etal_09}, which has an IP of 97eV and is a clear AGN
indicator. Five galaxies (NGC520, NGC253, NGC1614, NGC4676 and
NGC7252) which do not have a confirmed AGN signature so far, are
located in the lower left corner of the plot, where the two branches
merge. Since at this area of the plot classification is more
ambiguous, it is possible that a weak AGN component could be present,
even though its [NeV] emission is too weak to be detected.

Unfortunately, in most of the ULIRGs with available IRS high
resolution spectroscopy, their [SIII]33.48$\mu$m emission lines are
redshifted out of the wavelength coverage of the IRS
\citep[see][]{far_etal_07}. Among the 10 ULIRGs that have both
[OIV]25.89$\mu$m and [SIII]33.48$\mu$m detected, 9 are located on the
upper branch and they all have detected [NeV] emission. IRAS09022-3615
is the only ULIRG with no [NeV] detection and it is located between the
two branches. This is again consistent with the assertion that star
forming galaxies with an AGN component do follow the upper branch.

The lower branch of the diagram seems to be populated by sources in
which the ionizing radiation is due only to stellar photospheres. In
addition to the BCDs, the pure starburst galaxies NGC7714, NGC3310,
and NGC1222 \citep[e.g.][]{bra_etal_04,weh_etal_06, bec_etal_07} are
located in the lower branch before the two branches merge. Moving
along the branch towards higher values of the [NeIII]/[NeII] ratio,
the metallicity overall decreases (see \S 4). The harder radiation
field produced by the low metallicity environment strongly increases
the [NeIII]/[NeII] ratio (by up to a factor of 100), but increases
only moderately the [OIV]/[SIII] ratio (by a factor of 10). This was
also noticed by \citet{lut_etal_98}, who indicated that
low-metallicity dwarfs exhibit relatively stronger [OIV] than
low-excitation starbursts, but the effect is not very pronounced
compared to the large excitation difference as measured by the
[NeIII]/[NeII]. One galaxy with a detected [NeV] emission
\citep[NGC3628, see][]{ber_etal_09} is on the lower branch. We
note though that its location is also at the far left where the two
branches merge, and as a result it does not contradict our bifurcation
scheme.

\begin{figure}[ht]
\centerline{
\includegraphics[angle=0.,width=\hsize]{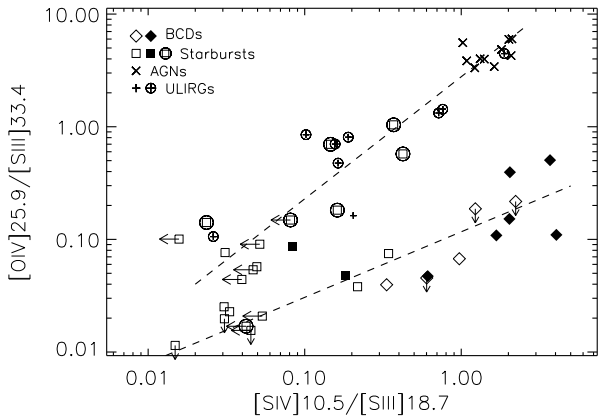}}
\caption{A plot of the [OIV]25.89$\mu$m/[SIII]33.48$\mu$m line ratio as
 a function of the [SIV]10.51$\mu$m/[SIII]18.71$\mu$m for our samples.
 The symbols are the same as  in Figure~\ref{os3_neratio}. Note that 
the bifurcation into two branches, indicated again with the dashed lines, 
is also visible. 
 \label{os3_sratio}}
\end{figure}

As with the [NeIII]/[NeII] ratio, the ratio of
[SIV]10.51$\mu$m/[SIII]18.71$\mu$m has also been widely used as a
probe of the hardness of the ionization field. When plotting the
[OIV]25.89$\mu$m/[SIII]33.48$\mu$m as a function of the [SIV]/[SIII]
line ratio, we observe a similar bifurcation signature (see Figure
~\ref{os3_sratio}). The sources located on the upper and lower
branches in Figure~\ref{os3_sratio} are the same to the ones found on
the upper and lower branches in Figure~\ref{os3_neratio}. The
separation of the two branches though is not as clear as in
Figure~\ref{os3_neratio}, mainly because of the narrower range of the
[SIV]/[SIII] values for our sample than the range of the
[NeIII]/[NeII] ratios. In addition, in several starbursts (8 out of
the 24 galaxies) no [SIV] emission was detected, thus making a clear
definition of the merging points of the two branches challenging.


\subsection{[OIV] Emission as an AGN Diagnostic}

\subsubsection{The [OIV]/[SIII] ratio and optical classification}

\begin{figure}[ht]
\centerline{
\includegraphics[angle=0.,width=\hsize]{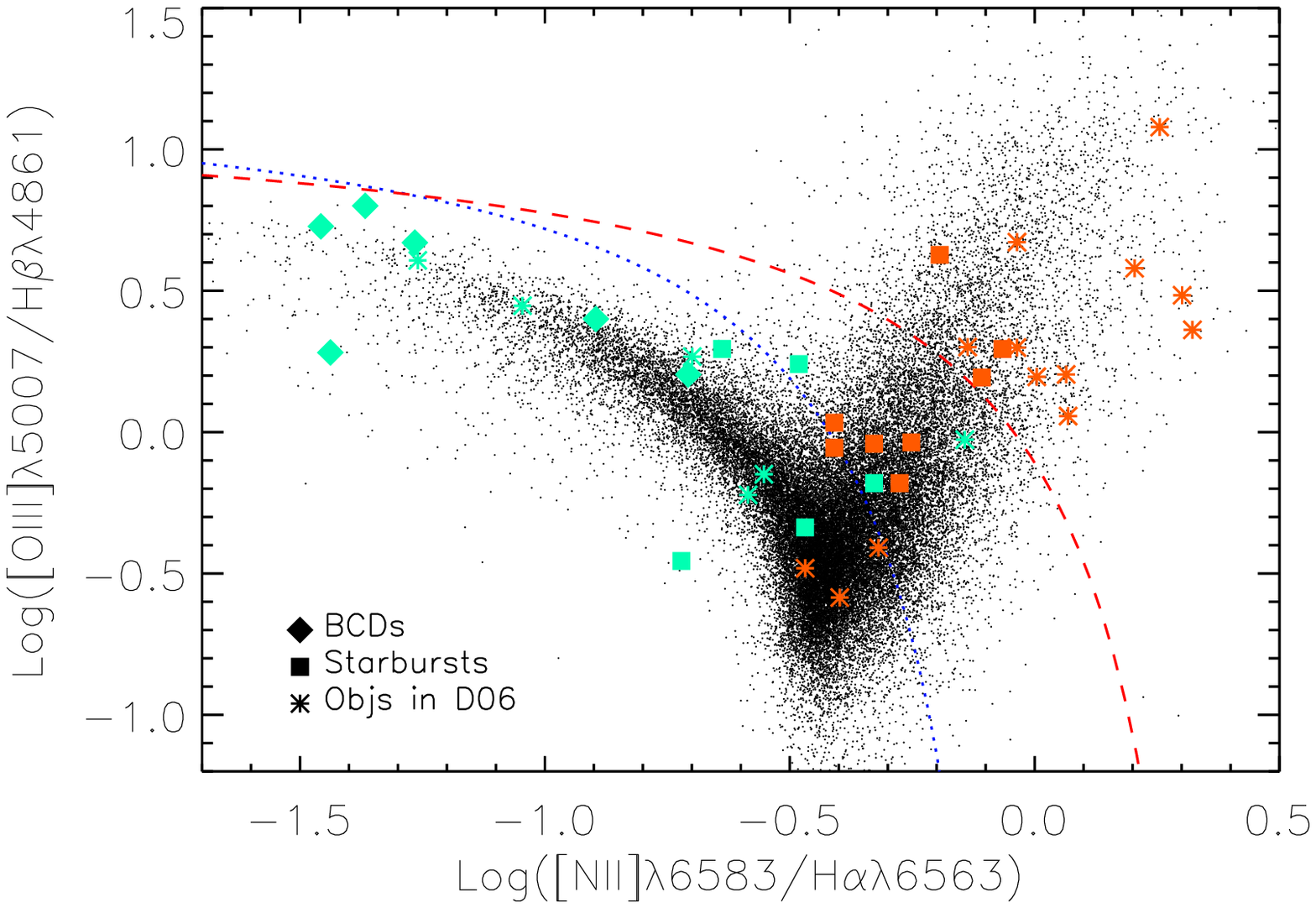}}
\caption{ The optical log([NII]/H$\alpha$) vs. log([OIII]/H$\beta$)
  diagram in which we include $\sim$40,000 SDSS galaxies, marked as
  black dots. The red dashed line and the blue dotted line are taken
  from \citet{kew_etal_01} and \citet{kau_etal_03} and they are used
  to separate AGNs, to the top right of the plot, from starburst
  galaxies to the bottom left. We include on the diagram our sample of
  BCDs and starbursts as well as the star-forming galaxies from
  \citet{dal_etal_06} (D06). Note that some galaxies in our starburst sample and in D06 also host an active nucleus and as a result are found in the AGN locus.  Galaxies located at the upper and lower branch
  of Figure ~\ref{ne32_o4s3_dale} are plotted in orange and cyan
  respectively (see text).
  \label{diagram}}
\end{figure}

Close examination of the [OIV]/[SIII] vs. [NeIII]/[NeII] diagram
presented in the previous section suggests that it can be used as an
AGN-starburst diagnostic, similar to the log([NII]/H$\alpha$)
vs. log([OIII]/H$\beta$) diagram \citep[e.g. the BPT
diagram,][]{bal_etal_81}, which has been used traditionally to
distinguish AGNs from starbursts in the optical \citep{vei_ost_87,
  kew_etal_01, kau_etal_03}. In both diagrams, star forming galaxies
are distributed along two separate branches, each having one end
respectively anchored by the location of AGNs and BCDs. In
Figure~\ref{diagram}, we plot on the BPT diagram $\sim$40,000 galaxies
from the Sloan Digital Sky Survey \citep[SDSS,
see][]{hao_etal_05_agn1}. Strong AGNs have high [NII]/H$\alpha$ and
[OIII]/H$\beta$ ratios and they are found at the tip of the right
branch.  Low-metallicity BCDs have high [OIII]/H$\beta$ but low
[NII]/H$\alpha$ ratios and they are located at the tip of the left
branch. The natural separation of galaxies into two branches indicates
again a different excitation mechanism. The blue dotted line,
empirically defined by \citet{kau_etal_03}, separates the two
branches. These authors classify galaxies below the line as star
forming galaxies, and those above it as AGNs. The red dashed line is
taken from \citet{kew_etal_01}, and demarcates the maximum position
that can be obtained by pure photo-ionization models. Galaxies located
above the line require an additional excitation mechanism, such as an
AGN, or strong shocks. The general classification scheme using the two
separation lines \citep{kew_etal_06} considers objects above Kewley's
line as AGN dominated sources, between Kewley's and Kauffmann's line
as composite AGN and starburst sources, and below the Kauffmann's line
as star forming galaxies.

To quantitatively check the analogy between our [OIV] diagram
(Figure~\ref{os3_neratio}) and the log([NII]/H$\alpha$)
vs. log([OIII]/H$\beta$) diagram used in the optical, we have
collected from the literature existing optical data for our starbursts
and BCDs. Special care was taken to match the optical apertures to the
11'' wide \spitzer IRS-LH slit, in which the [OIV] line is
located. This is especially important for the starburst galaxies as
many of them are extended. The most appropriate optical datasets for
the majority of galaxies were the integrated spectra obtained by
\citet{mou_ken_06} from which we retrieved the optical spectra data
for 13 starbursts and 8 BCDs.

\begin{figure}[ht]
\centerline{
\includegraphics[angle=0.,width=\hsize]{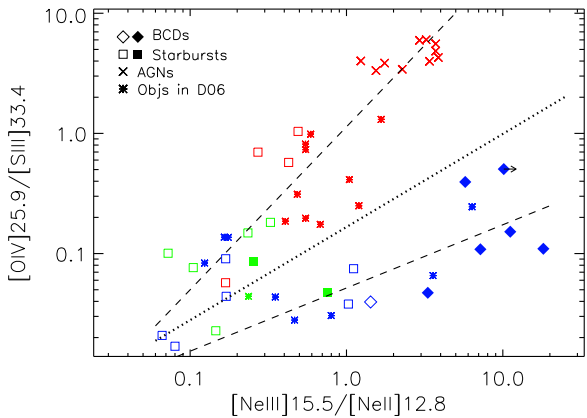}}
\caption{This is the same plot to the one we present in Figure 1, but
  we include also the star-forming galaxies from \cite{dal_etal_06}
  (D06, marked as stars). As before, filled symbols are galaxies with WR
  characteristics in the optical. The color of each symbol is based on
  the to the optical classification of the galaxy from
  Figure~\ref{diagram}. Galaxies above Kewley's line in
  Figure~\ref{diagram} are red, between Kewley's and Kauffmann's line
  are green, and below Kauffmann's line are blue. The dashed lines are the same as in
  Figure 1. The dotted line is Equation (1), defined empirically as
  the separatrix between the two branches.
  \label{ne32_o4s3_dale}}
\end{figure}

To expand our dataset, we also use the data for star-forming galaxies
from \citet{dal_etal_06}, which have matchable \spitzer
high-resolution mapping spectra extracted with an aperture of
23''$\times$15'', and optical spectra taken with an aperture of
20''$\times$20''. There are 19 objects with both detections in
[OIV]25.89$\mu$m and optical lines used in the BPT diagram, which are
displayed in the Figure 2 of \citet{dal_etal_06}.  In our
Figures~\ref{diagram} and~\ref{ne32_o4s3_dale}, we place our samples
with optical data, as well as the star-forming galaxies from
\citet{dal_etal_06}, in the corresponding optical log([OIII]/H$\beta$)
vs. log([NII]/H$\alpha$) and infrared [OIV]/[SIII] vs. [NeIII]/[NeII]
diagrams.The colors of the data symbols in each diagram are assigned
by their corresponding classification in the other diagram (see Figure
captions). In Figure~\ref{ne32_o4s3_dale}, the galaxies found above
Kewley's line in Figure~\ref{diagram} are red{\footnote{We do not have the matching optical data for the AGNs in our sample, but they are expected to lie well to the right of the Kewley's line.}}, between Kewley's and
Kauffmann's line are green, and below Kauffmann's line
are blue. We found that most optically identified
strong AGNs (marked with red $\times$) and starbursts (blue squares) are well separated in the upper and lower
branches, especially for values of the [NeIII]/[NeII] ratio greater
than $0.3$. There are four galaxies optically classified as starbursts
(blue squares) located in the upper branch in
Figure~\ref{ne32_o4s3_dale}. However, we note that their location in
the optical diagram in Figure~\ref{diagram} is near the locus where
the two branches merge, an area where classification of individual
objects is ambiguous.

In the optical diagram of Figure~\ref{diagram}, the data points are
colored based on their classification in the infrared diagram of
Figure~\ref{ne32_o4s3_dale}. More specifically, we empirically define
a line separating the upper and lower branch, having the functional
form of:

\begin{equation}
log(\rm{[OIV]/[SIII]})=0.08\times log(\rm{[NeIII]/[NeII]})+0.014
\end{equation}

Galaxies located on the upper branch above the line are plotted in
Figure~\ref{diagram} in orange, and below the line in cyan. This
demonstrates that the upper and lower branch of the infrared diagram
correspond well to the right and left branch of the optical.

We should stress once more that the optical spectra are collected from
various sources and have been obtained with different apertures. This
likely introduces uncertainties in the separatrix mentioned above,
which cannot be easily quantified. Even for datasets where optical and
infrared apertures are matched, wavelength dependent extinction will
result in discrepancies between the different classifications, in
particular for dusty sources, since the infrared can probe into
regions which are optically thick. Therefore, it is not surprising to
have some outliers that do not correspond to the same regions in both
diagrams. For example, galaxies located between Kewley's and
Kauffmann's line (green objects in
Figure~\ref{ne32_o4s3_dale}) may lie either on the upper or lower
branch of the infrared diagram. Despite these uncertainties, the overall consistency of the
two diagrams is very encouraging, convincing us that the [OIV]/[SIII]
vs. [NeIII]/[NeII] diagnostic along with equation (1) can be used to
classify the energy source of nuclear emission in galaxies,
particularly for those with high dust content.
 
\subsubsection{The [OIV]/[SIII] ratio and other infrared diagnostics}

\begin{figure}[ht]
\centerline{
\includegraphics[angle=0.,width=\hsize]{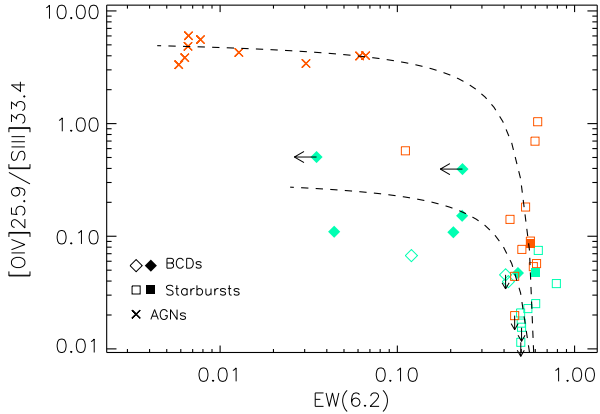}}
\caption{The [OIV]25.89$\mu$m/[SIII]33.48$\mu$m ratio as a function of the
6.2$\mu$m PAH EW. Starburst galaxies (squares), BCDs (diamonds) and
AGNs (crosses) located at the upper and lower branch of
Figure~\ref{ne32_o4s3_dale} are plotted in orange and cyan
respectively. Filled symbols are again WR galaxies. Two dashed lines
are drawn to indicate the trend of the upper and lower branches.
\label{eqpah_os3}}
\end{figure}

The ratio of the flux of [OIV]25.89$\mu$m line with the flux of lower-ionization line has been already proposed to diagnose the
dominant source of power in ULIRGs \citep{gen_etal_98,
  arm_etal_07}. In the Genzel et al. study, the ratio of
[OIV]25.89$\mu$m to the [NeII]12.81$\mu$m (or [SIII]33.48$\mu$m) was
plotted against the strength of the 7.7$\mu$m PAH. The PAH features
are strong in starbursts and weak in AGNs, as the PAH molecules are
easily destroyed by the hard radiation of the accretion disk or
outshined by the hot dust continuum emission from the AGN torus.  In
Figure~\ref{eqpah_os3}, we plot the [OIV]/[SIII] ratio as a function
of the equivalent width (EW) of the 6.2$\mu$m PAH feature. The PAH EWs
for the galaxies in our sample are collected from \citet{bra_etal_06,
  wu_etal_06}, and \citet{spo_etal_07}. As was the case with
Figure~\ref{diagram} colors of the galaxies in Figure~\ref{eqpah_os3}
are based on their mid-infrared classification in
Figure~\ref{ne32_o4s3_dale}. Those found on the upper branch above the
dotted separatrix line are marked in orange, while those on the lower
branch are in cyan. We find that the two color groups define two
separate trends, with the AGNs residing at the tip of one group and
the BCDs at the tip of the other. The upper trend is similar to the
trend found in the traditional Genzel et al.  diagram (indicated with
the upper dashed line), while the other trend clearly falls below
it. This suggests that the method to diagnose the fraction of AGN and
starburst power in the mid-IR for a given source suggested by
\citet{gen_etal_98}, can only be applied to sources which are part of
the upper trend in Figure~\ref{eqpah_os3}. The upper and lower dashed lines can be represented as equation (2) and (3) respectively:
\begin{equation}
log(\rm{[OIV]/[SIII]})-log(6.0)=0.19/(log(\rm{EW(6.2)})-log(0.7))
\end{equation}
\begin{equation}
log(\rm{[OIV]/[SIII]})-log(0.35)=0.51/(log(\rm{EW(6.2)})-log(0.7))
\end{equation}

\begin{figure}[ht]
\centerline{
\includegraphics[angle=0.,width=\hsize]{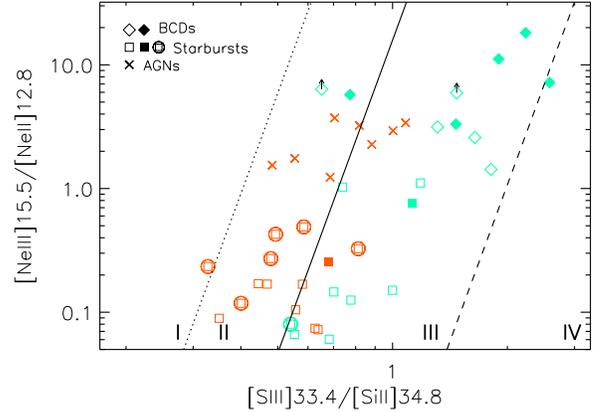}}
\caption{The diagram of [SIII]33.48$\mu$m/[SiII]34.81$\mu$m as a
  function of [NeIII]15.56$\mu$m/[NeII]12.81$\mu$m proposed by Dale et
  al. (2006). The diagram is divided into four regions (I-IV). Galaxies
  located in regions I and II should be classified as AGN with a $1
  \sigma$ confidence interval of 83\%-97\% and 73\%-88\% respectively,
  and galaxies found in regions III and IV should be classified as
  star-forming systems with a $1 \sigma$ confidence interval of
  84\%-93\% and 91\%-98\%, respectively. The boundaries of the four
  regions are indicated by lines of the same slope but different
  offset: log([NeIII]/[NeII])=8.4log([SIII]/[SiII]+$\gamma$, where
  $\gamma$=($+$3.3,$+$1.2,$-$2.5) for the dotted, solid and dashed
  lines. We include the BCDs (diamonds), starbursts (squares) and AGN
  (crosses) of our sample in the diagram. The coloring scheme follows
  the one of Figures 3 and 5, so that galaxies located at the upper
  and lower branch of Figure ~\ref{ne32_o4s3_dale} are indicated here
  in orange and cyan respectively. Starbursts with a [NeV]14.32$\mu$m
  detection are indicated with a circle around their corresponding
  symbols. Filled symbols are WR galaxies.
\label{ssi_neratio}}
\end{figure}

\citet{dal_etal_06} have also proposed a diagnostic diagram based on
the [SIII]33.48$\mu$m/[SiII]34.81$\mu$m line ratios as a function of
the [NeIII]15.56$\mu$m/[NeII]12.81$\mu$m ratios to distinguish AGN
from starburst emission. Their method relies on the fact that [SiII],
having an IP of only 8.2 eV, is a low-ionization emission line and a
strong coolant of X-ray-irradiated gas \citep{mal_etal_96}. AGNs
usually have stronger [SiII] emission than starburst galaxies and
diagnostic diagrams involving [SiII] present an advantage in cases of
faint sources where [OIV] is not easily detected.  In
Figure~\ref{ssi_neratio}, we plot the galaxies from our sample on the
[SIII]/[SiII] vs [NeIII]/[NeII] diagram, similarly to the Figure 9 of
\citet{dal_etal_06}. The galaxies are divided into four regions, and
are classified accordingly in a statistical sense: galaxies located in
regions I and II should be classified as an AGN with a $1 \sigma$
confidence interval of 83\%-97\% and 73\%-88\% respectively, and
galaxies found in regions III and IV should be classified as a
star-forming systems with a $1 \sigma$ confidence interval of
84\%-93\% and 91\%-98\%, respectively. In Figure~\ref{ssi_neratio} ,
the galaxies are also colored the same way as in Figure~\ref{diagram}
and Figure~\ref{eqpah_os3}. We find that AGNs and star-forming
galaxies situated on the upper branch are preferentially located in
regions I and II, while BCDs and star forming galaxies on the lower
branch are found in regions III and IV. This suggests that the two
methods overall agree.

\subsection{The nature of the [OIV]25.89$\mu$m emission in starbursts}

As mentioned in the previous section, the [OIV]25.89$\mu$m emission in
galaxies cannot originate from the HII regions around main-sequence
stars. Instead, it has to be associated either with an AGN component,
or with WR stars (Crowther 1999) and/or shock excitation
\citep{lut_etal_98}.  We have already shown in \S 3.2 and \S 3.3 that
the [OIV] diagram can successfully identify galaxies where an AGN
contributes to their mid-IR spectrum as they are placed on the upper
branch of the Figure~\ref{os3_neratio}. In this section, we present a
preliminary analysis on the nature of the [OIV] emission in the
starburst galaxies and BCDs of the lower branch, focusing on the
interpretation of models which include WR stars and shocks. More
detailed study on this subject will be presented in an upcoming paper.

Some of the sources in our sample are classified as Wolf-Rayet
galaxies. They show features of WR stars in their optical spectra,
most commonly the broad HeII$\lambda4686\AA$ emission which has an IP
of $\sim 54$eV, similar to [OIV]. Therefore, identifying correlations
between WR features and [OIV] emission could, in principle, reveal the
contribution of those stars. We cross-matched our BCDs with the WR
galaxy catalog presented in \citet{schaer_etal_99} and found that
seven BCDs (IZw18, SBS0335-052E, UM461, IIZw40, NGC1569, Mrk1450, and
NGC1140) as well as two starbursts (NGC1614 and NGC7714) have reported
WR features.

Unfortunately, this WR galaxy sample is small, and there are no
consistent measurements of the optical WR features from the
literature. Thus a direct correlation of the WR strength and the [OIV]
line emission could not be established. In Figure~\ref{os3_neratio} we
have marked these WR galaxies \citep{schaer_etal_99} as filled
symbols. It appears that at least the detections of WR signatures are
not closely associated with high [OIV]/[SIII] ratios, instead, they
appear to be more correlated with high [NeIII]/[NeII] ratios.

Another way to probe the contribution of the WR stars to the [OIV]
emission is by photo-ionization modeling. We perform a tentative
exploration of the contribution of those stars to the presence of the
O$^{3+}$ ion in our lower branch galaxies with Starburst 99 (Leitherer
1999) and CLOUDY (Ferland 2001). We use Starburst 99 to model the
stellar radiation field of an instantaneous burst of star formation
aging from 1 to 10 Myrs. All stars have a metallicity of 1/3
Z$_{\odot}$ and the follow a Salpeter IMF. The output from Starburst
99 is used as input to CLOUDY in order to estimate the flux of the
mid-IR emission lines.

The models predict that the presence of considerable [OIV] emission is
directly related to the appearance of the Wolf-Rayet phases, becoming
strong around 4 Myrs after the onset of the burst and fading to
undetectable levels 2 Myrs later. On the other hand, the [NeIII] line
is related to the presence of massive main sequence O stars but also
becomes very weak after 6 Myrs. Both [NeII] and [SIII] lines are still
strong after 10 Myrs. Consequently the [OIV]/[SIII] and [NeIII]/[NeII]
line ratios peak around 4-5 Myrs after the onset of the burst with
values of $\sim$0.6 and 40 respectively. The value of 0.6 for the
[OIV]/[SIII] line ratio agrees with the highest value observed in the
galaxies of the lower branch in Figure~\ref{os3_neratio}, but the
predicted [NeIII]/[NeII] ratio is about a factor of 4 higher than the
observed value. Clearly a more systematic study exploring thoroughly
the parameter space of the models and comparing them to larger sample
of galaxies with mid-IR measurements is required to further
investigate this discrepancy.

\begin{figure}[ht]
\centerline{
\includegraphics[angle=0.,width=\hsize]{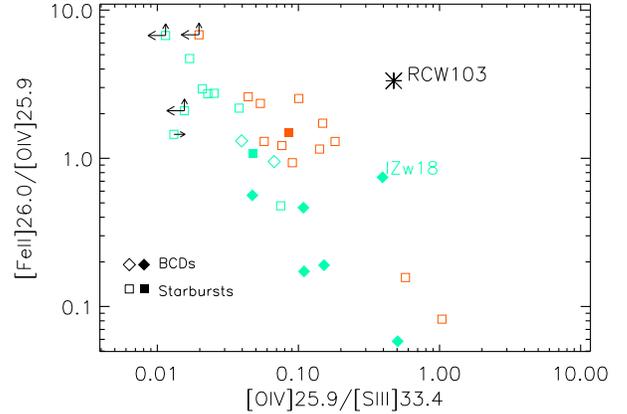}}
\caption{A plot of the [OIV]25.89$\mu$m/[SIII]33.48$\mu$m as a
  function of the [FeII]25.99$\mu$m/[OIV]25.89$\mu$m ratio for the BCDs
  (diamonds) and starbursts (squares). Galaxies located on the upper
  and lower branch of Figure ~\ref{ne32_o4s3_dale} are plotted in
  orange and cyan respectively. The large asterisk indicated the supernovae 
  remnant RCW03  \citep{oli_etal_99}, while the filled symbols are WR galaxies.
  \label{o4s3_fe2o4}}
\end{figure}

We also study the effects of shocks on the production of [OIV]
utilizing infrared lines which are strong in regions of shocks, such
as the [FeII]25.99$\mu$m \citep[see][]{lut_etal_03}. The infrared
[FeII] lines have been suggested to trace well of the supernova
activity \citep[][see also O'Halloran et al. 2008]{gre_etal_91}. The
[FeII] ionic abundances in supernova remnants are orders of magnitudes
larger than the typical values found in Orion-like HII regions
\citep[e.g.][]{sew_etal_83, gra_etal_87, oli_etal_89}. This is
primarily due to shock destruction of grains. Since iron is normally
depleted onto the grains, the abundance of iron in the ISM is
significantly increased in supernova remnants. Furthermore, the cooler
and denser gas behind a fast shock generates more low-ionized Fe$^+$
compared to what is found in typical HII regions. Following the
analysis of \citet{lut_etal_03} to evaluate the effects of shocks in
NGC6240, we plot in Figure~\ref{o4s3_fe2o4} the ratio of
[FeII]25.99$\mu$m/[OIV]25.89$\mu$m as a function of
[OIV]25.89$\mu$m/[SIII]33.48$\mu$m. Also plotted are the measurements
of the supernovae remnant RCW103 \citep{oli_etal_99}. Most BCDs and
starbursts in our sample show an anti-correlation on the diagram, with
no obvious indication of enhanced [FeII] emission. However, one
galaxy, IZw18, appears to be an outlier compared to other BCDs with a
clear enhancement of the [FeII] emission relative to the [OIV]
emission. This enhanced [FeII] emission could indicate a significant
contribution of shocks in IZw18.

The influence of shocks to the observed [OIV] emission can be better
constrained with detailed shock models \citep[e.g.][]{all_etal_08} but
we defer this work to a future paper on this topic.  Overall, our
preliminary study described in this section suggests that there is no
strong indication that shocks play an important role in the mid-IR
emission of starbursts and BCDs. Hence, WR stars are the most likely
culprit of the [OIV] emission seen in the lower-branch galaxies of the
Figure~\ref{os3_neratio}. We do note though the difficulty to
simultaneously fit the [OIV]/[SIII] and [NeIII]/[NeII] ratios with
current photo-ionization models. It appears that the models require
mechanisms to significantly suppress [NeIII]/[NeII] ratios for a fixed
[OIV] emission. This issue is discussed more in \S 4.

\subsection{The [NeV]14.32$\mu$m Emission} 

\begin{figure}[ht]
\centerline{
\includegraphics[angle=0.,width=\hsize]{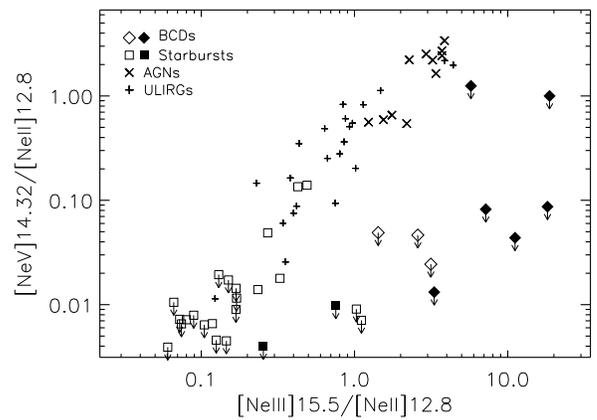}}
\caption{A plot of the [NeV]14.32$\mu$m/[NeII]12.81$\mu$m as a
  function of the [NeIII]15.56$\mu$m/[NeII]12.81$\mu$m ratio for BCDs,
  starbursts, AGNs, and ULIRGs. The symbols are the same as in
  Figure~\ref{os3_neratio}. Note that since no [NeV] was detected in
  any BCD only upper limits are indicated for the corresponding line flux ratios.
\label{ne52_neratio}}
\end{figure}

As we have discussed earlier, BCDs display a significantly harder
radiation field than typical starbursts with high [OIV]/[SIII] and
[NeIII]/[NeII] ratios. However, no [NeV]14.32$\mu$m emission has been
detected in any of them. It is known that [NeV], with an IP of 97eV,
serves as an unambiguous indicator of an AGN and the 7 out of our 24
starburst galaxies where [NeV]14.32$\mu$m was detected, all host an
AGN \citep[see table 1 of ][]{ber_etal_09}. In
Figure~\ref{ne52_neratio}, we plot the
[NeV]14.32$\mu$m/[NeII]12.81$\mu$m ratio as a function of
[NeIII]15.56$\mu$m/[NeII]12.81$\mu$m for BCDs, starbursts, AGNs, and
ULIRGs and find a similar bifurcation signature to the one in
Figure~\ref{os3_neratio}. This also confirms the bifurcation nature of
the [OIV] diagram discussed in \S 3.2 and \S 3.3. A similar
bifurcation is also seen in the [NeV]14.32$\mu$m/[NeII]12.81$\mu$m
vs. [SIV]10.512$\mu$m/[SIII]18.712$\mu$m diagram (not plotted).

\section{General Excitation of Star Forming Galaxies}

In addition to the high-ionization emission lines of [NeV] and [OIV],
other common mid-IR emission lines such as [SIV]10.51$\mu$m,
[SIII]18.71$\mu$m, [NeII]12.81$\mu$m, and [NeIII]15.56$\mu$m have been
used in order to probe the ionization field in BCDs and starbursts,
the [NeIII]15.56$\mu$m/[NeII]12.81$\mu$m vs.
[SIV]10.51$\mu$m/[SIII]18.71$\mu$m diagram being such an example
\citep[e.g.][]{ver_etal_03, dal_etal_06, far_etal_07}. Both
\citet{wu_etal_06} and Bernard-Salas et al. (2009) studied this
diagram independently for different samples. Here we combine both and
also include the data from our comparison sample of AGN, ULIRGs, and
HII regions. The study is similar to the one presented by
\citet{gro_etal_08}, who also include samples of various galaxy
types. The advantage of the present work is that the measurements are
more consistent as all data were obtained with the \spitzer
high-resolution modules.

\begin{figure}[ht]
\centerline{ \includegraphics[angle=0.,width=\hsize]{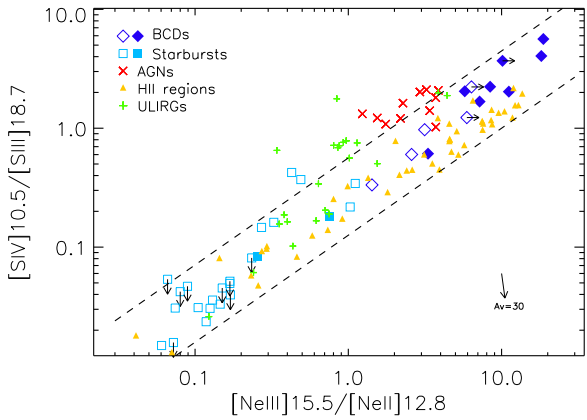}}
\caption{A plot of [NeIII]/[NeII] vs. [SIV]/[SIII] for BCDs,
  starbursts, AGNs, Galactic/LMC/SMC HII regions and ULIRGs. The
  symbols of BCDs, starbursts, AGNs, and ULIRGs are the same as in
  Figure~\ref{os3_neratio}. The HII regions are designated with
  triangles. Most BCDs, starburst galaxies, and HII regions are
    distributed in a narrow band, indicated with two dashed lines.\label{sratio_neratio}}
\end{figure}


In Figure \ref{sratio_neratio}, we plot the
[NeIII]15.56$\mu$m/[NeII]12.81$\mu$m and the
[SIV]10.51$\mu$m/[SIII]18.71$\mu$m ratios for the starbursts, BCDs,
AGNs, HII regions, and low-redshift ULIRGs. For the HII regions and
the ULIRG sample, only sources with detections in all four lines are
displayed. In addition, we include the SH mapping observations of the
central positions of NGC5253 \citep{bei_etal_06}, a typical BCD with
metallicity of 1/6 Z$_\odot$ \citep{kob_etal_99}. Since WR features
have also been found in the center of NGC5253, we designate it with a
filled diamond as other BCDs with detected WR features.

Observing the plot we note that starbursts populate the lower-left
corner of the plot \citep[see also][]{ber_etal_09}. BCDs in general
have higher values in both line ratios than the starbursts, while
those with WR signatures systematically have even higher
[NeIII]/[NeII] and [SIV]/[SIII] ratios than those without. The values
of [NeIII]/[NeII] in BCDs reach as high as 20, and [SIV]/[SIII] as
high as 10, values which in some cases are even greater than those
seen in AGNs. BCDs, starbursts as well as giant HII regions all lie in
a band stretching over 3 orders of magnitude in [NeIII]/[NeII] and
[SIV]/[SIII]. The band, denoted by the two dashed lines, has a slope
of $\sim0.9$, similar to the slope obtained by \citet{gro_etal_08}. As
discussed by \citet{gro_etal_08}, AGNs and ULIRGs are offset from the
band displaying slightly higher [SIV]/[SIII] ratios relative to their
[NeIII]/[NeII] ratios. This is very likely due to the harder radiation
field in AGNs which excites [SIV] more easily than [NeIII]. However,
we notice that the offset is not as significant as in the [OIV]
diagram. This indicates that the [NeIII]/[NeII] and [SIV]/[SIII]
ratios are not very sensitive to the excitation mechanisms, and cannot
be used as diagnostics of the central energy source in a galaxy
\citep{leb_etal_08}.

The higher [NeIII]/[NeII] and [SIV]/[SIII] ratios in BCDs compared to
average starburst galaxies indicate a harder radiation field in
BCDs. This had already been well established \citep[e.g.][and
references therein]{ver_etal_03, dal_etal_06}. We note, though, that in our sample of BCDs and starbursts there is little
overlap in the values of the [NeIII]/[NeII] and [SIV]/[SIII] ratios
with a separatrix at [NeIII]/[NeII]$\sim 1.2$ and [SIV]/[SIII]$\sim
0.4$. Since our BCDs have preferentially sub-solar
metallicities\footnote{Here we adopt
  (Ne/H)$_\odot$=1.2$\times$10$^{-4}$ from Feldman \& Widing (2003),
  (S/H)$_\odot$=1.4$\times$10$^{-5}$ from Asplund et al. (2005) and
  (O/H)$_\odot$=4.9$\times$10$^{-4}$ from Allende Prieto et
  al. (2001).} \citep{wu_etal_08}, while the galaxies in the starburst
sample are more metal rich \citep{ber_etal_09}, their separation
in the diagram suggests that metallicity is likely the dominant
parameter determining the [NeIII]/[NeII] and [SIV]/[SIII] ratios.

\begin{figure}[ht]
\centerline{ \includegraphics[angle=0.,width=\hsize]{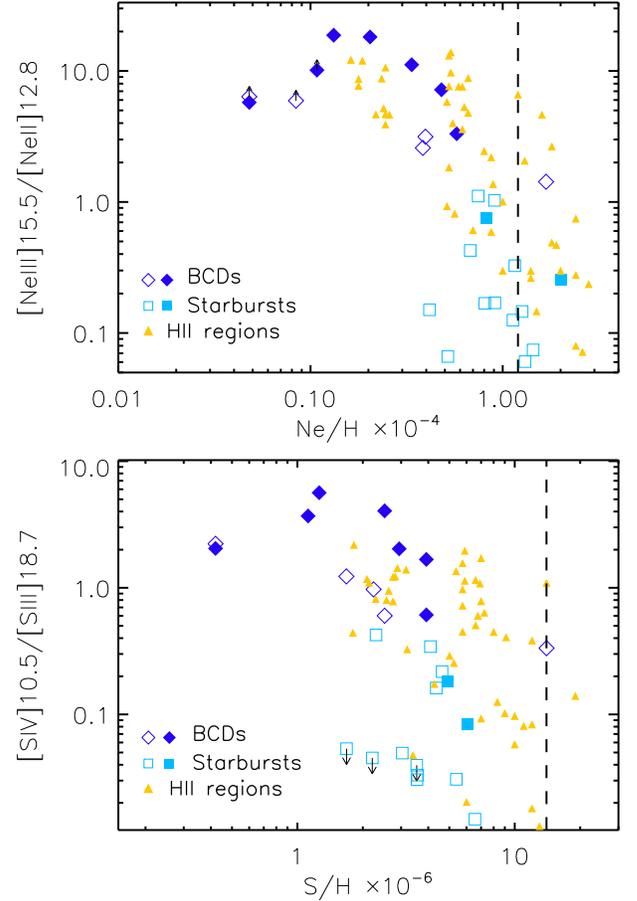}}
\caption{Top: A plot of the [NeIII]/[NeII] ratio as a function of the
  Ne abundance relative to hydrogen for BCDs, HII regions and
  starbursts with available abundance measurements. Bottom: Similar
  plot for the [SIV]/[SIII] ratio relative to the the S abundance. The
  object with the lowest measured Ne and S abundance is IZw18 and Tol
  65  \citep{wu_etal_08}. In both plots the vertical dashed line indicates the corresponding solar abundance (see text).
\label{abund_ne_neratio}}
\end{figure}
  
To examine this more directly, we use the metallicity measurements of
BCDs, HII regions, and starbursts from \citet{wu_etal_08},
\citet{leb_etal_08}, and Bernard-Salas et al. (2009), and plot in
Figure \ref{abund_ne_neratio}, the [NeIII]/[NeII] ratios vs. the Ne
abundances and [SIV]/[SIII] ratios vs. the S abundances. We observe an
anti-correlation between excitation and abundances for both Ne and S,
even though there is a substantial scatter \citep[see
also][]{ver_etal_03,wu_etal_06}.

As mentioned earlier, the low value of the [NeIII]/[NeII] ratio in
starbursts with high metallicities was noted already using ISO data
\citep[see also][]{tho_etal_00, ver_etal_03, mad_etal_06}, but its
implications are still puzzling. The low ratios apparently suggest a
lack of massive stars in the starbursts, in contradiction to the fact
that high-mass stars are clearly observed in nearby starbursts
\citep[e.g.][]{oco_etal_95, whi_sch_95,
  meu_etal_95}. \citet{tho_etal_00} attributed this to an aging
effect: massive starbursts have a shorter lifetime, thus, the hottest
stars may not be the dominant source of ionization in the
starbursts. However, \citet{rig_rie_04} propose that as high-mass
stars spend much of their lives embedded within ultra-compact HII
regions \citep[i.e.][]{cha_etal_08}, the high density and high
extinction of the ultra-compact HII regions prevents the mid-infrared
nebular lines from forming and escaping.

We note that the starbursts with low [NeIII]/[NeII] ratios have fairly
high [OIV]/[SIII] ratios (see Figure~\ref{os3_neratio}). This puts
additional constraints to the issue of low values of observed
[NeIII]/[NeII] ratios. As discussed in \S 3.4, if WR stars are the
main source of the [OIV] emission, the observed values of the
[OIV]/[SIII] in starbursts argue strongly against a low mass cutoff of
the IMF in these starbursts, since the WR phase would be too short to
sustain the intensity of the [OIV] observed. It will also places some
constraints on the scenario that massive stars spend much of their
life in ultra-compact HII regions. The strong extinction and high
density will have to somehow significantly affect [NeIII] and [NeII],
but not as much the [OIV] emission.

In Figure~\ref{abund_ne_neratio}, we notice that the two extremely low
metallicity objects, IZw18 and Tol65 \citep[see][]{wu_etal_08b}, deviate
from the anti-correlation of the emission line ratios vs. the
abundances. It appears as if the anti-correlation starts after a
maximum value for [NeIII]/[NeII] and [SIV]/[SIII] at $\sim$20 and
$\sim$8 respectively.  We should note though that extremely low metallicity galaxies are scarce, and their 
abundance measurements using the infrared fine-structure lines are
more challenging due to the non-detection of their Hu$\alpha$ 12.7$\mu$m
line. However, similar plots using the oxygen abundance determined
from optical spectroscopy also shows the same scatter
\citep{wu_etal_06}.

\section{Conclusions}

We have presented our analysis of the high spectral resolution mid-IR
observations for a sample of 12 BCDs obtained with Spitzer/IRS. A
direct comparison with samples of starburst galaxies, AGN, ULIRGs, as
well as isolated HII regions, enabled us to extend earlier studies of
the ionization and excitation in star forming galaxies towards lower
metallicities, and to use the observed mid-IR line emission to explore
the excitation mechanisms in these systems.

Although BCDs are known to have a hard radiation field, no
[NeV]14.32$\mu$m emission has been detected with Spitzer/IRS even for
the most metal-poor BCDs.

We find that the [OIV]25.89$\mu$m line is commonly detected in the
BCDs, starbursts and AGNs. We have demonstrated that the diagram of
[OIV]25.89$\mu$m/[SIII]33.48$\mu$m as a function of
[NeIII]15.56$\mu$m/[NeII]12.81$\mu$m can be useful diagnostic to probe
the dominant power source in dust enshrouded systems. Star forming
galaxies populate two branches in this diagram, with those harboring
an AGN component located on the upper branch, and those without on the
lower one. This diagram provides a classification consistent with the
log([NII]/H$\alpha$) vs. log([OIII]/H$\beta$) diagram in the optical,
especially for [NeIII]/[NeII] values above 0.3. We empirically define
a line (Equation 1) separating galaxies with and without AGN contribution, which
could be used in conjunction with other diagnostics to better
determine the presence of an AGN in the infrared. A similar
bifurcation is seen in a [NeV]/[NeII] vs. [NeIII]/[NeII] diagram. In the [NeIII]/[NeII] vs. [SIV]/[SIII] diagram. AGNs and ULIRGs are only
slightly off the band defined by BCDs, starbursts, and HII regions, with
higher [SIV]/[SIII] values for given [NeIII]/[NeII] ratios.

In an effort to probe the contribution of shock excitation to the
observed [OIV]25.89$\mu$m flux, we compare it with the
[FeII]25.99$\mu$m line , which is a good tracer of supernovae
activity. We found no evidence of shock contribution, demonstrated by
an enhanced [FeII] emission, in starbursts or BCDs. The only exception
is IZw18, which may indicate a significant shock contribution in this
low metallicity system.

The most likely contributor to the [OIV] flux, both in starbursts and
BCDs, are WR stars. However, current photo-ionization models predict
flux ratios of [NeIII]/[NeII] which are too high for the observed
[OIV]/[SIII] values. This places an additional constrain on the
problem of low [NeIII]/[NeII] ratios already known in high-metallicity
starbursts, and will be further investigated in future work.


\acknowledgments The
authors thank the anonymous referee for useful comments which greatly
improved the manuscript. We also thank John Moustakas for kindly providing us the optical data of the SINGS sample. Support for this work was provided by NASA through
Contract Number 1257184 issued by the Jet Propulsion Laboratory,
California Institute of Technology under NASA contract 1407. 
V.C. acknowledges partial support from the EU
ToK grant 39965 and FP7-REGPOT 206469.


\newcommand{\noopsort}[1]{}

\appendix
%
%

\begin{figure*} 
\begin{center}
\resizebox{\hsize}{!}{\includegraphics{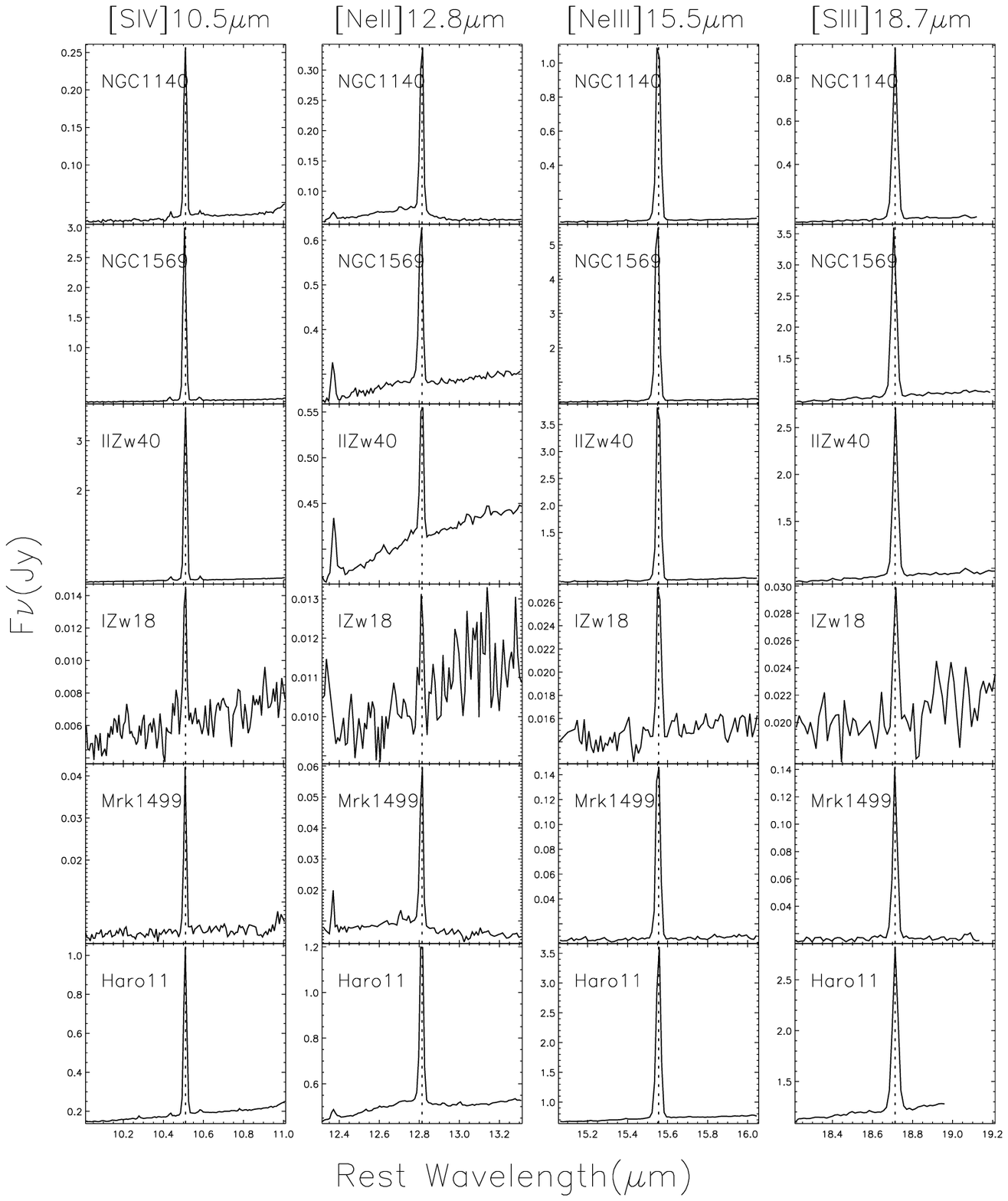}}
\caption{Zoom in of IRS high resolution spectra near the wavelength range of [SIV]10.51$\mu$m, [NeII]12.81$\mu$m, [NeIII]15.56$\mu$m, and [SIII]18.71$\mu$m for NGC1140, NGC1569, IIZw40, IZw18, Mrk1499, and Haro11. The expected location of the each emission line is indicated with the dashed vertical line.
  \label{lines1}}
\end{center}
\end{figure*}

\begin{figure*} 
\begin{center}
\resizebox{\hsize}{!}{\includegraphics{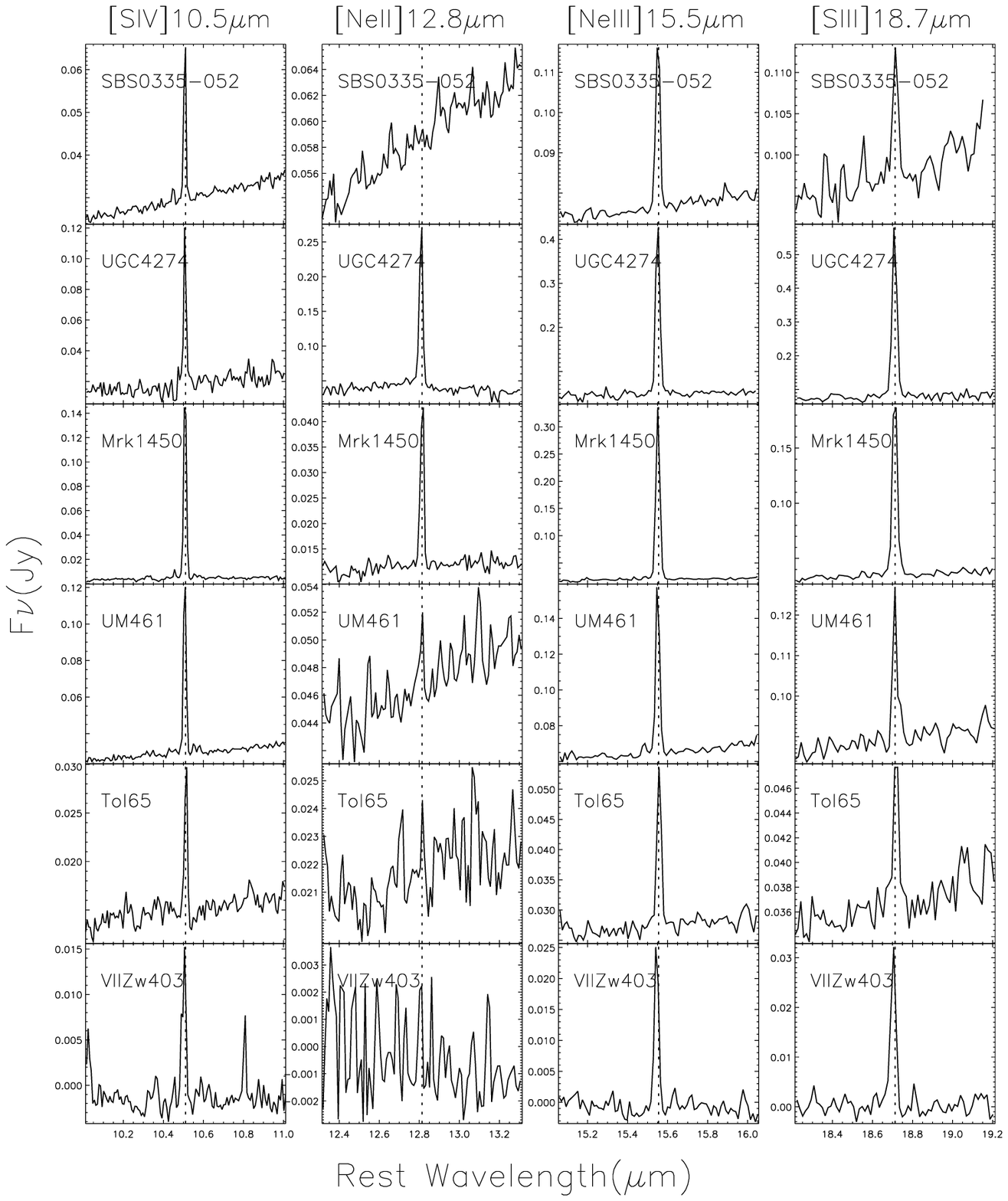}}
\caption{Zoom in of IRS high resolution spectra near the wavelength range of  [SIV]10.51$\mu$m, [NeII]12.81$\mu$m, [NeIII]15.56$\mu$m, and [SIII]18.71$\mu$m for SBS0335-052E, UGC4274, Mrk1450, UM461, Tol65, and VIIZw403.  The expected location of the each emission line is indicated with the dashed vertical line.
  \label{lines2}}
\end{center}
\end{figure*}

\begin{figure*} 
\begin{center}
\resizebox{\hsize}{!}{\includegraphics{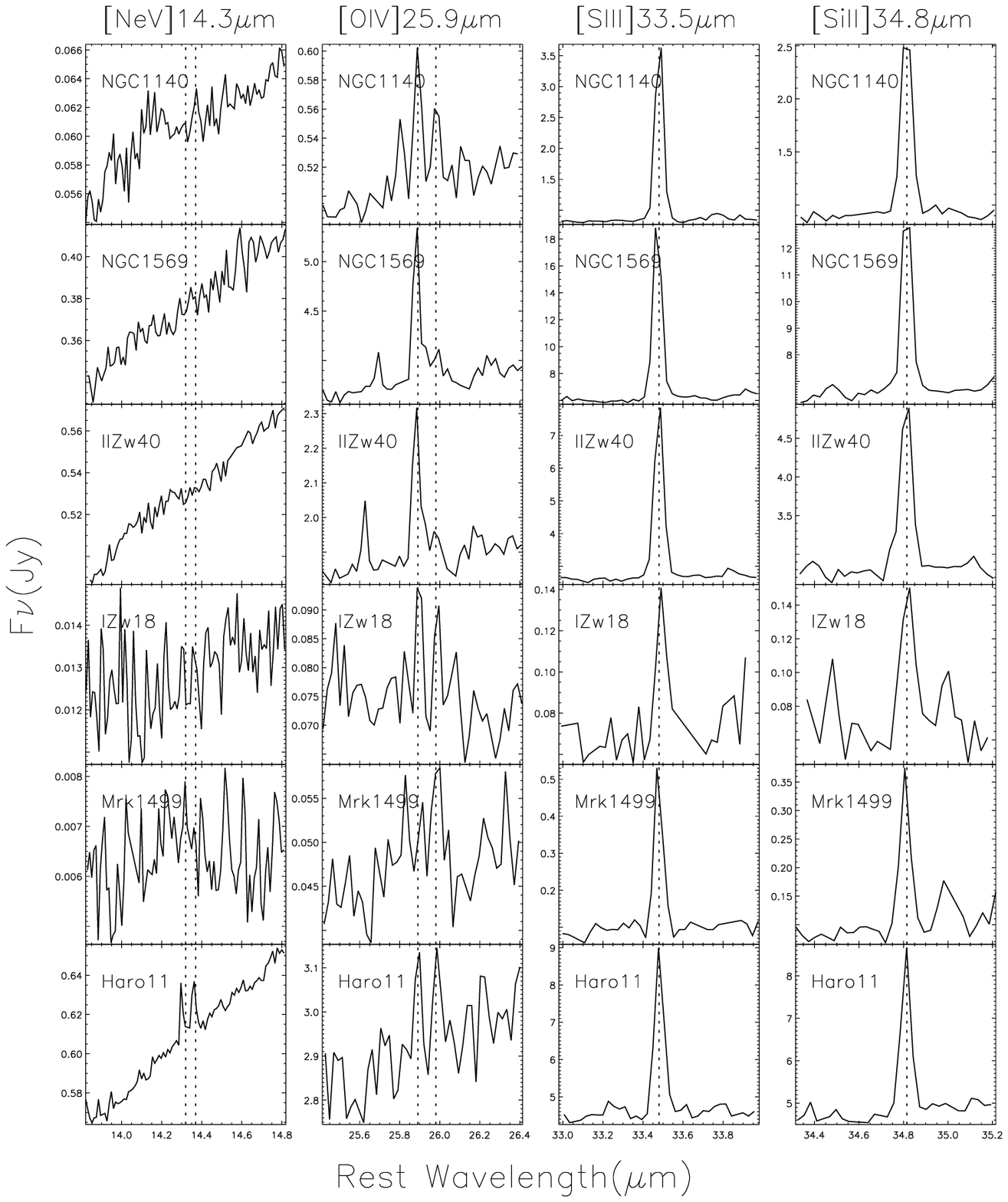}}
\caption{Zoom in of IRS high resolution spectra near the wavelength range of [NeV]14.32$\mu$m ( with its adjacent [Cl II]14.37$\mu$m), [OIV]25.89$\mu$m, [FeII]25.99$\mu$m, [SIII]33.48$\mu$m, and [SiII]34.81$\mu$m for NGC1140, NGC1569, IIZw40, IZw18, Mrk1499, and Haro11.  The expected location of the each emission line is indicated with the dashed vertical line.
  \label{lines3}}
\end{center}
\end{figure*}

\begin{figure*} 
\begin{center}
\resizebox{\hsize}{!}{\includegraphics{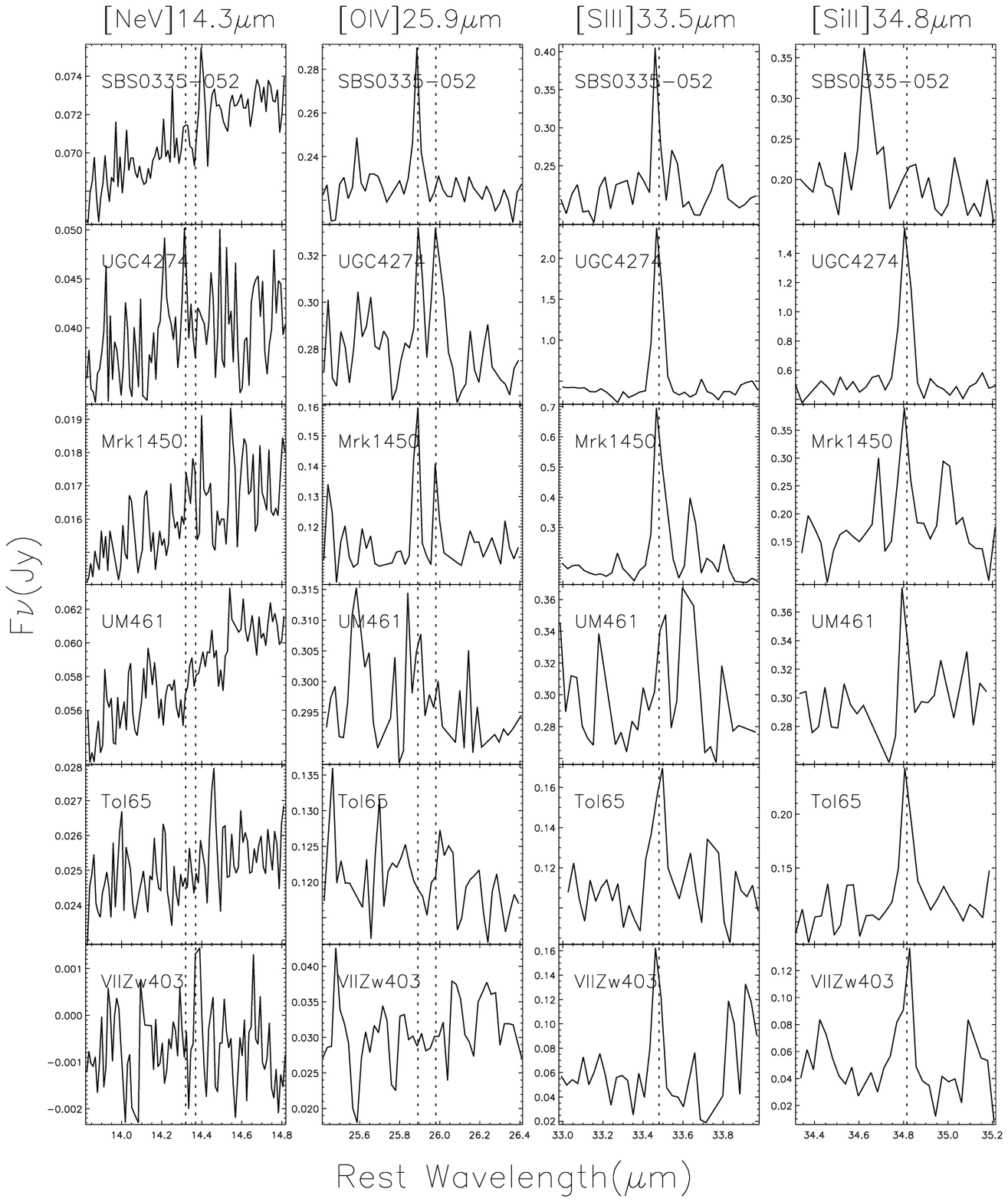}}
\caption{Zoom in of IRS high resolution spectra near the wavelength
  range of [NeV]14.32$\mu$m (with its adjacent [Cl II]14.37$\mu$m),
  [OIV]25.89$\mu$m, [FeII]25.99$\mu$m, [SIII]33.48$\mu$m, and
  [SiII]34.81$\mu$m for SBS0335-052E, UGC4274, Mrk1450, UM461,
  Tol1214-277, Tol65, and VIIZw403.  The expected location of the each
  emission line is indicated with the dashed vertical line.
\label{lines4}}
\end{center}
\end{figure*}


\clearpage
\begin{landscape}
\begin{deluxetable}{lrrrrrrrrrr}
  \tabletypesize{\scriptsize}
  \setlength{\tabcolsep}{0.05in}
  \tablecaption{Mid-Infrared Line Measurements of BCDs\label{tabline}}
  \tablewidth{0pc}
  \tablehead{
    \colhead{} & \multicolumn{10}{c}{Flux\tablenotemark{a} ($\times10^{-16}$W m$^{-2}$)}\\
    \colhead{Object Name} & \colhead{[SIV]} & \colhead{[NeII]} & \colhead{[NeIII]} & \colhead{[SIII]} & 
    \colhead{[SIII]} & \colhead{[SiII]} & \colhead{[NeV]} & \colhead{[NeV]} & \colhead{[OIV]} & 
    \colhead{[FeII]} \\
    \colhead{} & \colhead{10.51$\mu$m} & \colhead{12.81$\mu$m} & \colhead{15.56$\mu$m} &
    \colhead{18.71$\mu$m} & \colhead{33.48$\mu$m} & \colhead{34.81$\mu$m} & \colhead{14.32$\mu$m} & \colhead{24.32$\mu$m} & \colhead{25.89$\mu$m} &
    \colhead{25.99$\mu$m}\\
    \colhead{} & \colhead{34.8eV} & \colhead{21.6eV} & \colhead{41.0eV} & \colhead{23.3eV} & 
    \colhead{23.3eV} & \colhead{8.2eV} & \colhead{97.1eV} & \colhead{97.1eV} & \colhead{54.9eV} &
    \colhead{7.9eV} \\
  }
  \startdata  
  Haro11        & 4.396$\pm$0.030 &   3.114$\pm$0.027 & 9.801$\pm$0.058 &   4.523$\pm$0.041  &   6.773$\pm$0.351 &   5.165$\pm$0.267 &$<$0.076  &$<$0.554  &   0.456$\pm$0.116 &   0.433$\pm$0.188   \\
  NGC1140       & 1.226$\pm$0.014 &   1.138$\pm$0.025 & 3.770$\pm$0.032 &   2.007$\pm$0.013  &   4.071$\pm$0.078 &   2.777$\pm$0.088 &$<$0.015 &$<$0.081  &   0.192$\pm$0.037 &   0.108$\pm$0.044   \\
  SBS0335-052E   & 0.155$\pm$0.004 &   $<$0.014        & 0.142$\pm$0.005 &   0.042$\pm$0.009  &   0.204$\pm$0.027 &$<$0.150           &$<$0.012  &$<$0.042  &   0.103$\pm$0.018 &   0.006$\pm$0.007   \\
  NGC1569       &14.764$\pm$0.080 &   1.576$\pm$0.030 &17.583$\pm$0.095 &   7.270$\pm$0.055  &  19.197$\pm$0.468 &  10.113$\pm$0.444 &$<$0.069  &$<$0.460  &   2.917$\pm$0.223 &   0.554$\pm$0.124   \\
  IIZw40        &18.555$\pm$0.141 &   0.622$\pm$0.024 &11.299$\pm$0.085 &   4.581$\pm$0.045  &   8.199$\pm$0.156 &   3.672$\pm$0.215 &$<$0.054  &$<$0.286  &   0.899$\pm$0.119 &   0.155$\pm$0.066   \\
  UGC4274       & 0.422$\pm$0.016 &   0.898$\pm$0.028 & 1.281$\pm$0.030 &   1.265$\pm$0.033  &   3.008$\pm$0.156 &   1.661$\pm$0.124 &$<$0.044  &$<$0.100  &   0.119$\pm$0.032 &   0.156$\pm$0.037   \\  
  IZw18         & 0.045$\pm$0.007 &   0.008$\pm$0.002 & 0.046$\pm$0.005 &   0.022$\pm$0.005  &   0.109$\pm$0.024 &   0.141$\pm$0.049 &$<$0.010  &$<$0.035  &   0.043$\pm$0.008 &   0.032$\pm$0.010   \\
  VIIZw403      & 0.107$\pm$0.013 &   $<$0.015       & 0.089$\pm$0.006 &   0.087$\pm$0.007  &   0.159$\pm$0.074 &   0.108$\pm$0.038 &$<$0.015 &$<$0.040 &        $<$0.030  &        $<$0.030    \\
  Mrk1450       & 0.734$\pm$0.010 &   0.134$\pm$0.007 & 0.963$\pm$0.008 &   0.438$\pm$0.010  &   0.774$\pm$0.157 &   0.300$\pm$0.107 &$<$0.011  &$<$0.040  &   0.084$\pm$0.011 &   0.039$\pm$0.010   \\
  UM461         & 0.456$\pm$0.009 &   0.015$\pm$0.005 & 0.281$\pm$0.008 &   0.081$\pm$0.006  &   $<$0.153        &   0.091$\pm$0.018 &$<$0.015  &$<$0.064  &         $<$0.045  &          $<$0.045   \\
  Tol65         & 0.069$\pm$0.007 &   $<$0.014        & 0.089$\pm$0.005 &   0.031$\pm$0.006  &   0.129$\pm$0.037 &   0.198$\pm$0.037 &$<$0.009  &$<$0.024  &          $<$0.028 &           $<$0.028  \\
  Mrk1499       & 0.175$\pm$0.006 &   0.194$\pm$0.009 & 0.502$\pm$0.006 &   0.291$\pm$0.005  &   0.616$\pm$0.039 &   0.375$\pm$0.046 &$<$0.009  &$<$0.025  &         $<$0.028  &          $<$0.028   \\
  \enddata
  \tablenotetext{a}{The integrated fluxes are measured from the high
    resolution spectra of the targets.  Background emission has not
    been subtracted and no scaling between the SH and LH modules of
    IRS has been applied. The central wavelength and the ionization
    potential of each line are indicated.}
  
\end{deluxetable}
\clearpage
\end{landscape}
\end{document}